\documentclass[fleqn]{mnras}
\usepackage{graphicx}
\usepackage[utf8]{inputenc}
\usepackage{bm}

\usepackage{natbib}
\usepackage{newtxtext,newtxmath}
\usepackage[T1]{fontenc}
\usepackage{ae,aecompl}

\title[Solid Phosphorus and Fluorine in 67P/C-G]{The Detection of Solid Phosphorus and Fluorine in the Dust from the
Coma of Comet 67P/Churyumov-Gerasimenko}

\author[E. Gardner et al.]{Esko Gardner$^{1}$\thanks{Contact e-mail:
    \href{mailto:esgard@utu.fi}{esgard@utu.fi}}, Harry J. Lehto$^{1}$,
  Kirsi Lehto$^2$, Nicolas Fray$^3$, Anaïs Bardyn$^4$
  \newauthor{Tuomas Lönnberg$^5$, Sihane Merouane$^6$, Robin Isnard$^3$, Hervé Cottin$^3$,}
  \newauthor{Martin Hilchenbach$^6$ \& the COSIMA team}
  \\
$^1$Tuorla Observatory, Department of Physics and Astronomy, 20014 University of Turku, Finland \\
$^2$Molecular Plant Biology, Department of Biochemistry, 20014 University of Turku, Finland \\
$^3$Laboratoire Interuniversitaire des Systèmes Atmosphériques (LISA),
UMR CNRS 7583, Université Paris-Est-Créteil, Université de Paris, \\Institut Pierre Simon Laplace (IPSL), Créteil, France \\
$^4$Earth and Planets Laboratory, Carnegie Institution of Washington, 5241 Broad Branch Road, Washington, DC 20015, USA\\
$^5$Department of Chemistry, 20014 University of Turku, Finland \\
$^6$Max-Planck-Institut für Sonnensystemforschung, Justus-von-Liebig-Weg 3, 37077 Göttingen, Germany}

\date{Updated: \today}
\pubyear{2020}

\begin{document}
\label{firstpage}
\pagerange{\pageref{firstpage}--\pageref{lastpage}}
\maketitle
\begin{abstract}
Here we report the detection of phosphorus and fluorine
in solid particles collected from the inner coma of comet
67P/Churyumov-Gerasimenko measured with the COSIMA instrument on-board
the Rosetta spacecraft, only a few kilometers away from the comet nucleus. We have detected phosphorus-containing
minerals from the presented COSIMA mass spectra, and can rule out e.g. apatite minerals as the
source of phosphorus. This
result completes the detection of life-necessary CHNOPS-elements in
solid cometary matter, indicating
cometary delivery as a potential source of these elements to the
young Earth. Fluorine was also detected with CF$^+$ secondary ions originating from the cometary dust.

\end{abstract}


\section{Introduction}

Comets are remnants from the Solar protoplanetary disc
\citep{Willacy2015}. Being formed beyond the ice-line orbiting the Sun, on
average, at distances further than the asteroid belt, and experiencing less
processing, they are thought to represent the most pristine matter of the
Solar System.

The first detection of phosphorus in a comet came over 30 years ago
from the report by \cite{Kissel1987} with a
line at m/z=31 in the PUMA mass spectrometer.  This was from the summed spectra from cometary dust collected during flyby of Comet 1P/Halley by
the Vega 1 mission in 1986. The interpretation was that no molecule could have
survived the impact during dust capture at a velocity of $>$ 70 kms$^{-1}$, and thus this line could only be attributed to atomic
phosphorus (Kissel, private communication). It is unknown in what kind of parent mineral this
phosphorus was contained.

The second detection of phosphorus has been reported in dust particles collected by the NASA Stardust spacecraft during the flyby of comet 81P/Wild in 2004, and returned to Earth in 2006 \citep{Flynn2006,Joswiak2012}. It was further
analyzed by \cite{Rotundi2014}, where phosphorus was detected in a
single cometary particle and associated with the presence of calcium. They concluded that
phosphorus was most likely contained within an apatite particle
\citep{Rotundi2014}. However, studies of the nanoscale mineralogy of Wild 2 also suggested phosphide minerals as phosphorus carriers \citep{Joswiak2012}.

Another detection of phosphorus and fluorine came from the ROSINA DFMS
instrument on board Rosetta \citep{Altwegg2016,Dhooghe2017,Rivilla2020}. In these cases, they detected
elemental phosphorus, PO and CF in the gas phase of 67P/Churyumov-Gerasimenko (67P/C-G).

We report here the detection phosphorus and fluorine in mass spectra measured from solid
dust particles of 67P/C-G.
Previous studies have already reported detection of C, H, N and O in the dust particles of comet 67P/C-G \citep{Fray2016,Fray2017,Bardyn2017,Paquette2018,Isnard2019}. It is also estimated that the organic component of the dust particles is made of high molecular weight material \citep{Fray2016}, that represents about 45\% in mass of the dust particles \citep{Bardyn2017}.   
S has been well detected showing a strong signal in the mass spectra
\citep{Paquette2017,Bardyn2017}.

In the process of forming life, water soluble reactive phosphorus
compounds were required to convert nucleotide precursors by
phosphorylation to active nucleotides. 

Reduced phosphorus minerals, such as schreibersite (Fe,Ni)$_3$P
\citep{Pasek2017b}, could have been available on early Earth both from
meteoritic, and very hot volcanic sources
\citep{Pasek2017a,Britvin2015,Turner2018}. However, unlike the
other elements required for life (CHNOS), gaseous forms of phosphorus
were unlikely to have been present as a major species in the early
Earth atmosphere, and thus was required to be in solid and soluble
form \citep{Pasek2017b,Pasek2019}.

\section{Methods} 
The COmetary Secondary Ion Mass Analyser (COSIMA), designed in the late 1990s and launched on-board Rosetta in 2004, is a Time-of-Flight
Secondary Ion Mass Spectrometer (TOF-SIMS) with a mass resolution of about
$m/{\Delta m}$=1400 at m/z=100 on board the Rosetta spacecraft, which
accompanied the comet 67P/Churyumov-Gerasimenko from August 2014
to September 2016 \citep{Kissel2007,Glassmeier2007,Hilchenbach2016}.  
During this time period COSIMA collected particles that originate from 67P/C-G
\citep{Langevin2016,Merouane2017}, at low impact
velocity ($< 10$ kms$^{-1}$) \citep{Rotundi2015} on silver and gold substrates. The grand total number of dust
particle fragments collected by COSIMA is more than 35,000 from an
estimated 1200-1600 original particles \citep{Merouane2017}, which
were fractured upon impact or in the subsequent collisions inside the
instrument. 

A beam of primary $^{115}$In$^+$ ions accelerated to 8 keV impacts the sample and releases secondary ions from the top surface of the particle or substrate \citep{Kissel2007}. The m/z of these secondary ions is measured by the time of flight spectrometer. The temperature inside the COSIMA instrument is about 283 K \citep{Bardyn2017}. We suppose that the interior of the instrument is in equilibrium with outside gas pressure of about 5$\times$10$^{-11}$ mbar, as measured by the instrument COPS \citep{Hoang2017}, the pressure changes due to heliocentric
 distance, latitude and location, but is correct to an order of magnitude. This is practically a vacuum, so the volatiles on the surface of the particles are lost between collection and measurement. The particles were stored between a few days and up to a year and a half before measurement, giving ample time for volatiles to escape. The instrument has two modes,
 positive and negative, sensitive to positive and negative ions,
 respectively.

Mass spectra were obtained from particles collected on 21 substrates, but due to limited time and resources, only about 250 particles have been analysed by TOF-SIMS. Most of the particles were given a name, to ease discussion about specific particles, and very small particles were numbered.

The instrument has a known contaminant, polydimethylsiloxane (PDMS), with significant peaks at m/z=73.05, in positive mode and m/z=74.99 is negative mode. These correspond to the ion fragments: Si(CH$_3$)$_3^+$ and CH$_3$SiO$_2^-$, respectively.

For the fitting process of spectra we have used a Levenberg–Marquardt
fit, fitting up to four peaks at each integer mass \citep{Stenzel2017}. The method
does not have a fixed peak list, but attempts to search for the
combination, which best fits the overall shape.


\section{The detection of phosphorus and fluorine in particles collected by COSIMA} 
For the purpose of this study, a set of summed spectra from 24 selected particles were analysed,
comparing them to a sample of a nearby background set. A background
set is a reference location on a substrate close to a particle, where there is
no visible cometary matter. Summing spectra allows elements with lower yield to be detected. For example, P$^+$ is particularly challenging for TOF-SIMS, as it yields signals over an order of magnitude lower than Fe$^+$ \citep{Stephan2001}.

The main focus in this study was to find various ionic species of
phosphorus present in TOF-SIMS spectra. We look for PO$_X^-$ in
negative spectra, ionic P
in negative and positive spectra and any other phosphorus associated compounds.

PH$_3^+$ (phosphine) and PH$_4^+$ (hydrogenated phosphine) were absent
from any analysed individual and summed positive spectra, which was
expected due to their volatile nature. This is the
same result as obtained by the ROSINA instrument in the gas phase, where PH$_3^+$ was not reliably detected \citep{Altwegg2016}. They
also did not find any indication of a parent mineral for phosphorus, but attribute it to be from a PO molecule \citep{Rivilla2020}. 

We noticed that phosphorus (P), mono-isotopic at m/z=30.97, is
detected when we sum a large amount of spectra. However, the signal is
too weak when viewed spectrum by spectrum. To aid detection we summed
the positive spectra acquired from a given particle. Out of the tested
24 particle sets, we found a significant contribution of phosphorus in
comparison to a local background in four particles: Uli, Vihtori,
Günter and Fred (see supplementary materials for more details). Fred
and Uli (shown in Figure \ref{PositiveP}), as well as Vihtori and Günter (see supplementary material), show clear cometary signals for CF$^+$ (m/z=30.9984) as well as P$^+$. This marks the first detection of both
CF$^+$ and P$^+$ in solid cometary dust. The detection of CF$^+$ originating from the dust particles complements the previous detection of F$^+$ presented in \cite{Dhooghe2017}. We searched for the signal of PO$^{2-}$ and PO$^{3-}$ in the cometary particles, but as the background spectra present a quite high signal of PO$^{2-}$ and PO$^{3-}$, there was no clear contribution of the cometary particles to these ions to be found.

\begin{figure*}

       \includegraphics[scale=0.75]{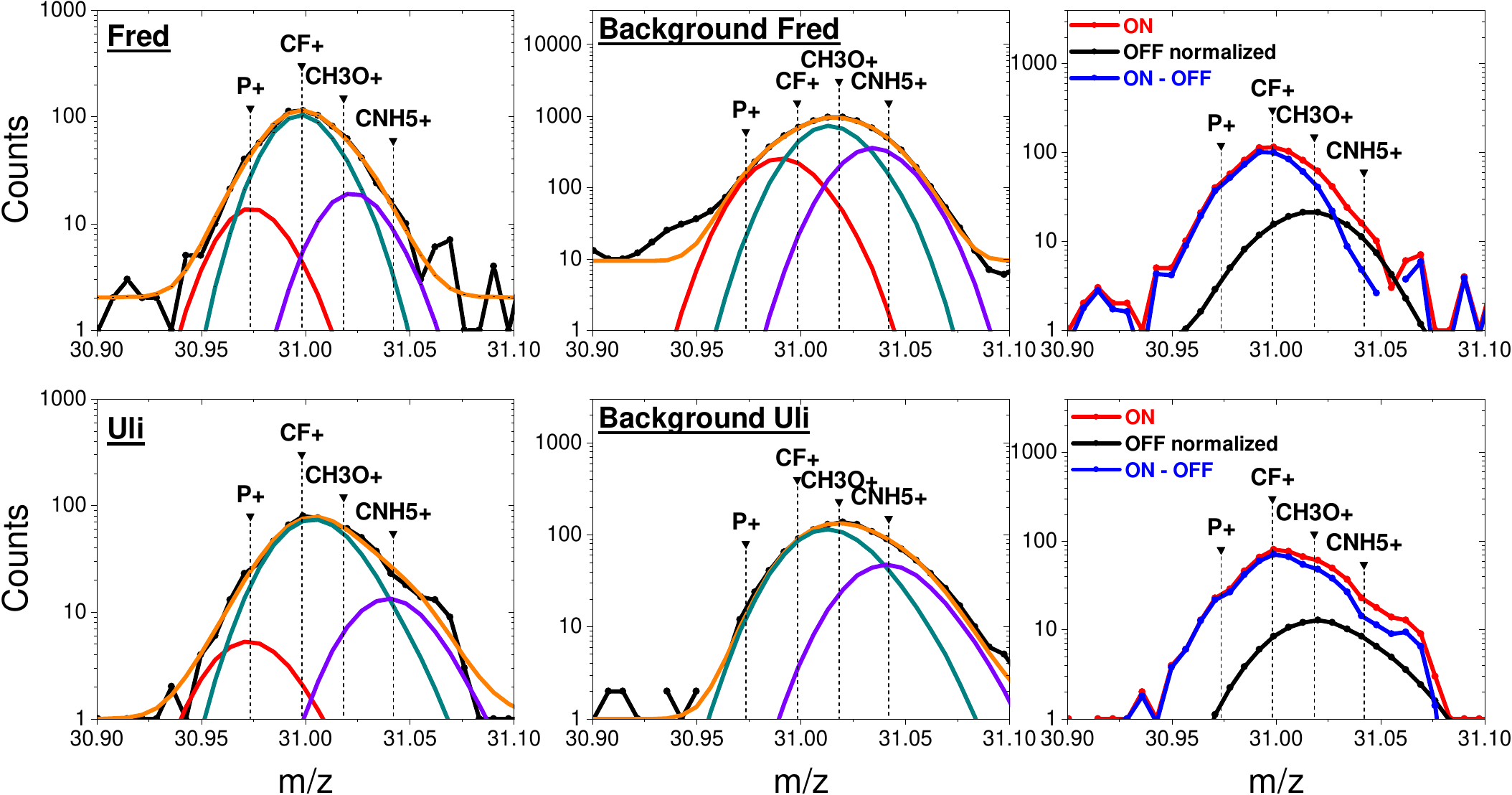}

\caption{Summed positive spectra (black line) for particle Fred (top left, 17 spectra) and Uli (bottom left, 10 spectra), and their comparative background sets (black line in the middle column, 5 and 2 spectra, respectively). The plots show the individual fits (red, cyan and purple) of multiple ions and the overall fit (orange line).The positions of the m/z of P$^+$, CF$^+$, CH$_3$O$^+$ and CNH$_5^+$ are shown in all panels to guide the eye. The right column shows the subtraction between the spectra on the particles and the normalized respective background spectra, where red shows the sum of the selected spectra taken on the particle and black shows the sum of the selected spectra taken on the target (next to the particle and at the same date), which has been normalised to the intensity of the PDMS fragment at m/z = 73.05. The spectra taken on the particles present a shift toward the left compared to the spectra acquired on the target (contamination). Thus at m/z = 31, the cometary contribution is located on the left side of the peak (the cometary contribution should have a negative mass defect at m/z = 31) which is an argument in favour of a contribution of the cometary particles to the signal attributed to P$^+$ and CF$^+$. The red and cyan individual fits (left column) are attributed to P$^+$ and CF$^+$, respectively. The errors on the position of these fits are less than one TOF channel which is of the same order than the difference between the positions of these fits and the exact mass of P$^+$ and CF$^+$.}\label{PositiveP}
\end{figure*}

\section{Comparison to reference samples} 
Using our reference COSIMA instrument at Max Planck Institute für
Sonnensystemforschung (MPS), Göttingen, Germany, we measured and
analysed two reference samples, fluoroapatite and schreibersite. Both
of these belong to families known to be found in meteorites
\citep{Hazen2013}. The first reference sample, fluoroapatite,
Ca$_5$(PO$_4$)$_3$F contains oxidized phosphorus. Apatite was chosen for its
terrestrial availability and its presence in meteorites. It was
commercially purchased and sourced from Cerro de Mercado,
Durango, Mexico. The apatite shows a clear signature of Ca$^+$,
P$^\pm$, PO$_2^-$ and PO$_3^-$. We do not see calcium in significant
amounts in these cometary dust samples, so the phosphorus likely cannot
be explained by apatite-like minerals. See table \ref{P-Comparison}
for a comparison of the cometary particles' yield of calcium and iron
in comparison to the reference samples. For all the particles on which P has been detected, the Ca$^+$/P$^+$  ionic ratio is much smaller than on apatite (Table \ref{P-Comparison}). Thus, we can rule out the apatite as the source of phosphorus.

\begin{table*}
\caption{Ionic ratios of $^{40}$Ca$^+$/ P$^+$ and $^{56}$Fe$^+$ / P$^+$. In all the cases, both of the ratios are much lower on the cometary particles than on the reference samples of apatite and schreibersite, which allows to rule out, at a significant level, the presence of Apatite, and possibly schreibersite in the cometary particles. Thus, the main carrier of phosphorus remains unknown. The errors are calculated from the Poisson error for the fitted lines, and are equivalent to 1-$\sigma$ errors.}\label{P-Comparison}
\begin{center}
\begin{tabular}[htbp]{lcc}
   \hline
   Substrate \& particle name & $^{40}$Ca$^+$/P$^+$ & $^{56}$Fe$^+$/P$^+$\\
   \hline
   1CF/Uli & 8.57 $\pm$ 1.93  & 38.17 $\pm$ 7.46 \\
   2CF/Vihtori & 1.20 $\pm$ 0.15   & 7.75 $\pm$ 0.70  \\
   2CF/Fred & 2.95 $\pm$ 0.53  & 20.24 $\pm$ 2.83  \\
   1D2/Günter & 0.74 $\pm$ 0.09  & 6.17 $\pm$ 0.49 \\
   \hline
   Apatite & 1332.5 $\pm$ 218.1  & N/A  \\
   Schreibersite & N/A & 428.4 $\pm$ 15.0  \\
   \hline

\end{tabular}
\end{center}
\end{table*}

Another reference sample was from the Fe-Ni-P phosphide group, schreibersite (Fe,Ni)$_3$P with reduced phosphorus. The schreibersite sample, of unknown (meteoritic, terrestrial or otherwise) origin, was obtained from the Mineral Sciences at Smithsonian Museum of Natural History. It shows a clear signature of P \& Fe. Our cometary samples show these, with a lower Fe to P ratio (Table \ref{P-Comparison}), which cannot confirm that the source of phosphorus is schreibersite. This is expected, if the phosphorus containing area is smaller than the beam size of the COSIMA primary beam (35 x 50$\mu$m$^2$).

Our conclusion is that here, the measured phosphorus is not from
apatite. The ion ratios for schreibersite also do not fit our findings. Phosphorus must come from another source, such as elemental phosphorus, or some other non-calcium containing mineral, although, as previously mentioned, it is probably not a phosphate because we could not find a clear cometary contribution of PO$^{2-}$ and PO$^{3-}$. Also, this means that the source of the fluorine is not from a fluorapatite.

\section{Discussion}
One of the challenges in understanding the origin of life processes, is the lack of soluble
phosphorus containing molecules in the terrestrial environments
\citep{Yamagata1991}. 

It has been experimentally shown that soluble P, HCN and H$_2$S can
serve as suitable feed stock for the prebiotic synthesis of
nucleotides, amino acids and phosphoglycerine backbones
\citep{Patel2015,Stairs2017}.

These reactions could be driven most efficiently by highly reduced
phosphorus, e.g. different mineral phosphides, such as those
belonging to the iron-nickel-phosphide group, known to occur mostly in
meteoritic materials
\citep{Gull2015,Pasek2005,Bryant2006,Herschy2018} or possibly elemental phosphorus. The phosphite anion
(PO$_3^{3-}$ or, given the conditions, HPO$_3^{2-}$) is a soluble and
highly reactive molecule, and is readily formed e.g. by the
hydrolysis of schreibersite \citep{Gull2015}.

So far, the different organic materials and feed-stocks regarding the
origin of life have been suggested to be derived either from 
meteoritic or geochemical origins
(\citealt{Kurosawa2013,Bada2013,Patel2015,Britvin2015}, and references
therein). However, the detection of all the life promoting compounds,
i.e.\ various CH compounds \citep{Fray2016,Isnard2019}, N \citep{Fray2017}, O
\citep{Paquette2018}, S \citep{Paquette2017}, and here, the solid forms
of P, in comet 67P/C-G, means that we have detected, in solid form, many ingredients regarded as important in current theories about origin of life. It is conceivable that
early cometary impacts onto the planet surface have been less energetic, as
compared to the impacts of the heavy stony meteorites \citep{Morbidelli2015}, thus preserving
the prebiotic molecules in a more intact condition.
\cite{Clark1988,Clark2018} suggested primeval procreative comet ponds
as the possible environment for the origin of life, while \cite{Chatterjee2016} suggested that hydrothermal impact craters
in icy environments could create another suitable cradle for life. In both cases phosphorus could be delivered by comets.

The results here indicate that elements for life can originate from solid cometary matter. It is possible to seed the required elements with solid cometary matter, that is rich in
volatiles. Although, more importantly, the compounds must be reactive and soluble, no matter how they are delivered. The solubility of the detected cometary phosphorus from 67P/C-G is not clear, but we can conclude that it cannot be Apatite, which is a common mineral source of phosphorus in meteorites. Additionally, other phosphate minerals are unlikely, because we could not find a clear cometary contribution of PO$^{2-}$ and PO$^{3-}$.

The presence of all the CHNOPS-elements give a strong premise for a future cometary sample-return mission to a comet. This could confirm the presence of all compounds and their possible mineral sources and the possible solubility of the matter. This would also allow for a comprehensive analysis of the relative amounts of these CHNOPS-elements.

\section*{Acknowledgements}
COSIMA was built by a consortium led by the
Max-Planck-Institut für Extraterrestrische Physik, Garching, Germany, in collaboration with Laboratoire de Physique et Chimie
de l’Environnement, Orléans, France, Institut d’Astrophysique
Spatiale, CNRS/INSU and Université Paris Sud, Orsay, France,
the Finnish Meteorological Institute, Helsinki, Finland, Universität Wuppertal, Wuppertal, Germany, von Hoerner und Sulger
GmbH, Schwetzingen, Germany, Universität der Bundeswehr, Neubiberg, Germany, Institut für Physik, Forschungszentrum Seibersdorf, Seibersdorf, Austria, and Institut für Weltraumforschung,
Österreichische Akademie der Wissenschaften, Graz, Austria, and
is lead by the Max-Planck-Institut für Sonnensystemforschung,
Göttingen, Germany. The support of the national funding agencies
of Germany (DLR), France (CNES), Austria and Finland and the
ESA Technical Directorate is gratefully acknowledged. We thank
the Rosetta Science Ground Segment at ESAC, the Rosetta Mission
Operations Centre at ESOC and the Rosetta Project at ESTEC for
their outstanding work enabling the science return of the Rosetta
Mission.\\

H. J. Lehto, E. Gardner and K. Lehto acknowledge the support of the
Academy of Finland (grant number 277375).\\

We acknowledge the Mineral Sciences at Smithsonian Museum of Natural
History for providing the schreibersite.\\

\section*{Data Availability}
The data underlying this article are available in the Planetary
Science Archive of ESA https://www.cosmos.esa.int/web/psa/psa-introduction, and in the
Planetary Data System archive of NASA https://pds.nasa.gov/.


\newcommand{\newblock}{}
\bibliographystyle{mnras}
\bibliography{biblio}
\bsp

\label{lastpage}
\end{document}


\begin{table*}
    \caption{Summary of particle information, including collection substrate, collection time
      and particle type as well as the amount of spectra summed for
      the particle. The particle type classification and volume is
      based on previous work (Langevin et al., 2016), except for Günter, which is calculated based on a
      half-sphere shape. The presence of phosphorus was identified for the first four listed particles.}\label{Particle-properties}
\begin{tabular}[htbp]{lccccc}
   \hline
   Substrate \&  & Collection & Particle & Particle volume & Number of\\
   particle name &  period  & type       & (10$^6$ $\mu$m$^3$) &spectra\\
   \hline
   1CF/Uli & 2014-12-20 $-$ 2014-12-27 & Compact & 1.66 &
   10\\
   2CF/Vihtori & 2015-01-24 $-$ 2015-01-25 & Glued
   cluster & 2.86 &21\\
   2CF/Fred & 2015-01-24 $-$ 2015-01-25 & Rubble
   pile / glued cluster & 12.8 & 17\\
   1D2/Günter & 2016-02-29  $-$ 2016-03-01 & Rubble
   pile & 13.0 &37\\
   \hline
\end{tabular}
\end{table*}

\begin{table*}
    \caption{Summary of particles analyzed by not found to have
      phosphorus, including collection substrate, collection time 
      and particle type as well as the amount of spectra summed for
      the particle. The particle type classification and volume is
      based on previous work (Langevin et al., 2016). Particles
      absent from that study have their volume
      calculated assuming they have a
      half-sphere shape.}\label{Particle-properties2}
\begin{tabular}[htbp]{lccccc}
   \hline
   Substrate \&  & Collection & Particle & Particle volume & Number of\\
   particle name &  period  & type       & (10$^6$ $\mu$m$^3$) &spectra\\
   \hline
  1CF/Alicia   & 2014-12-16 $-$ 2014-12-20 & Rubble pile & 1.67 & 14\\
  1CF/André    & 2015-01-09 $-$ 2015-01-14 & Glued cluster & 1.57 & 4\\
  1CF/Elly     & 2015-01-24 $-$ 2015-01-25 & Rubble pile & 0.43 & 6\\
  1CF/Hase     & 2015-01-24 $-$ 2015-01-25 & Rubble pile& 0.82 & 8\\
  2CF/Jean-Baptiste &  2015-01-24 $-$ 2015-01-25 & Glued / shattered cluster & 10.3 & 36\\
  2CF/Jean-Pierre & 2015-01-24 $-$ 2015-01-25 & Shattered cluster& 3.8 & 2\\
  2CF/Lari     & 2015-01-24 $-$ 2015-01-25 & Shattered cluster & 1.08 & 2\\
  3CF/Matt     & 2015-01-24 $-$ 2015-01-25 & Rubble pile& 2.24 & 12\\
  2CF/Jessica  & 2015-01-26 $-$ 2015-01-27 & Shattered cluster& 9.28 & 12\\
  1CF/Isbert   & 2015-01-28 $-$ 2015-01-29 & Rubble pile& 1.68 & 2\\
  3C7/Juvenal  & 2015-03-01 $-$ 2015-03-09 & Compact & 1.35 & 14\\
  2D1/Kenneth  & 2015-05-11 $-$ 2015-05-12 & Glued cluster & 1.75 & 8\\
  3D1/Sachi    & 2015-05-11 $-$ 2015-05-12 & Compact &  0.14 & 4\\
  1CD/Umeka    & 2015-07-03 $-$ 2015-07-04 & Rubble pile & 0.24 & 4\\
  2CD/Karen    & 2015-07-25 $-$ 2015-07-26 & Shattered cluster & 4.83 & 18\\
  1CD/Bonin    & 2015-07-31 $-$ 2015-08-01 & Shattered cluster & 1.17 & 8\\
  1CD/Devoll   & 2015-07-31 $-$ 2015-08-01 & Glued cluster & 0.14 & 2\\
  1D2/Juliette & 2015-10-23 $-$ 2015-10-29 & Glued cluster & 0.68 & 12\\
  1D2/Fadil    & 2015-11-16 $-$ 2015-11-18 & Rubble pile & 0.31 & 2\\
  2D2/Stefane  & 2016-01-17 $-$ 2016-01-18 & Rubble pile & 0.21 & 2\\   
   \hline
\end{tabular}
\end{table*}

\begin{figure*}
\includegraphics[scale=0.75]{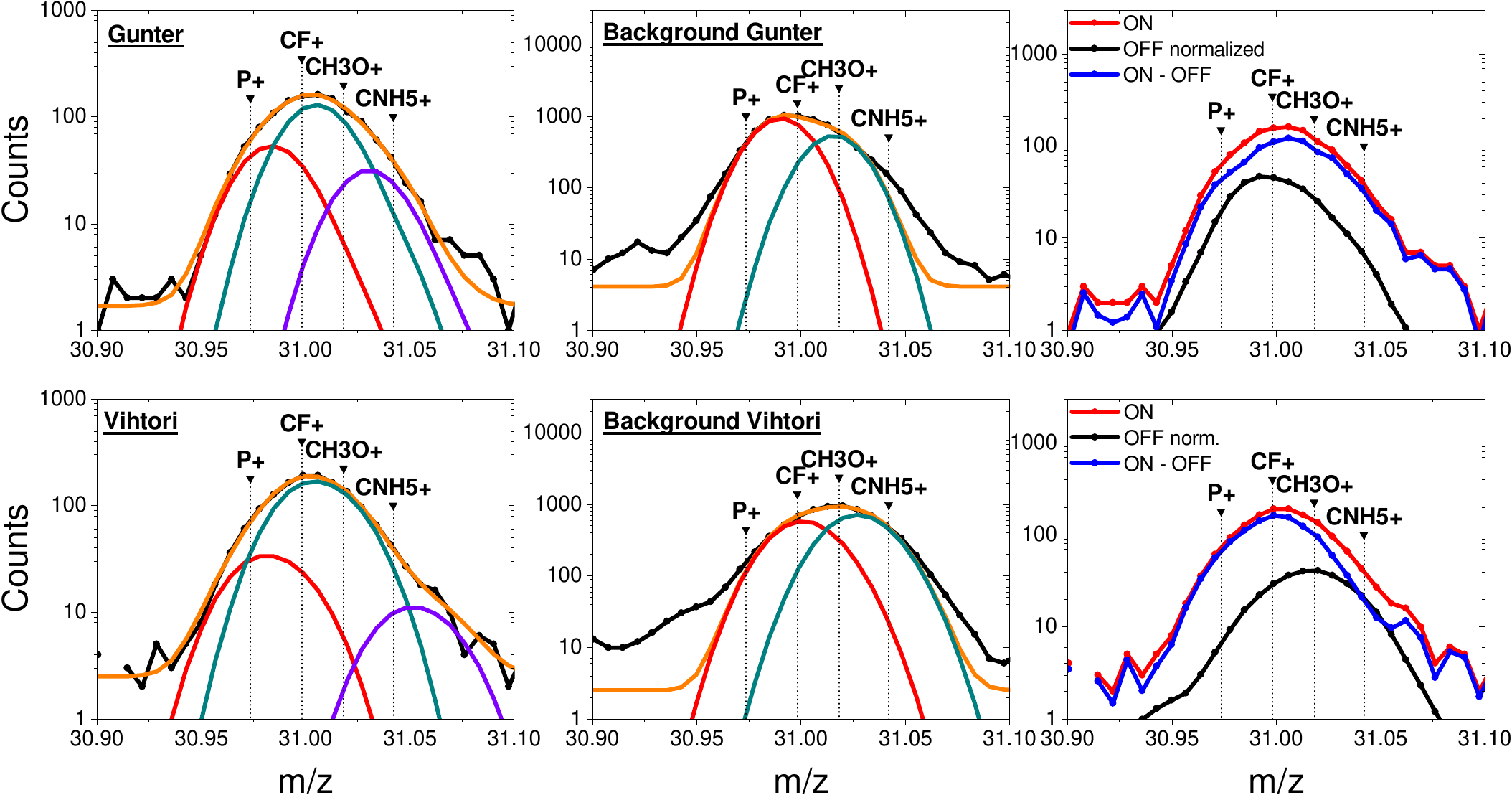}
\caption{As Fig 1 for particles Vihtori and Günter.}\label{VihtoriGunter}
\end{figure*}
\begin{figure*}
\includegraphics[scale=0.75]{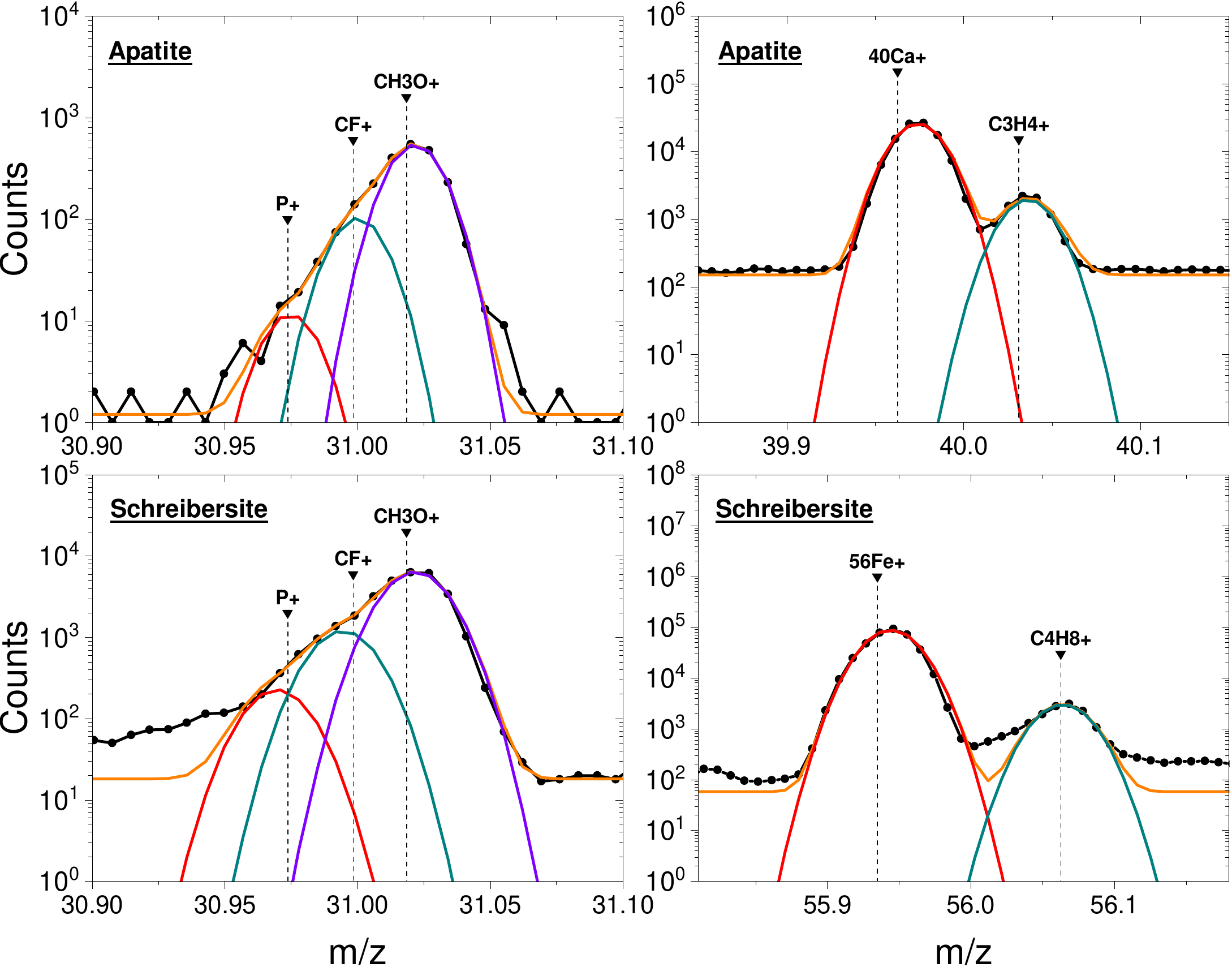}
\caption{Positive spectra for the reference materials measured on Earth on the reference COSIMA instrument, showing the presence of phosphorus as well as the main positive peak present in the substance. The mass calibration for Apatite is off at higher masses due to the methodology not being perfect for pure reference material. Notice the multiple orders of magnitude that separate the signal of Fe/Ca in comparison to P.}\label{reference-positive-spectra}
\end{figure*}
\begin{figure*}
            \includegraphics[scale=0.75]{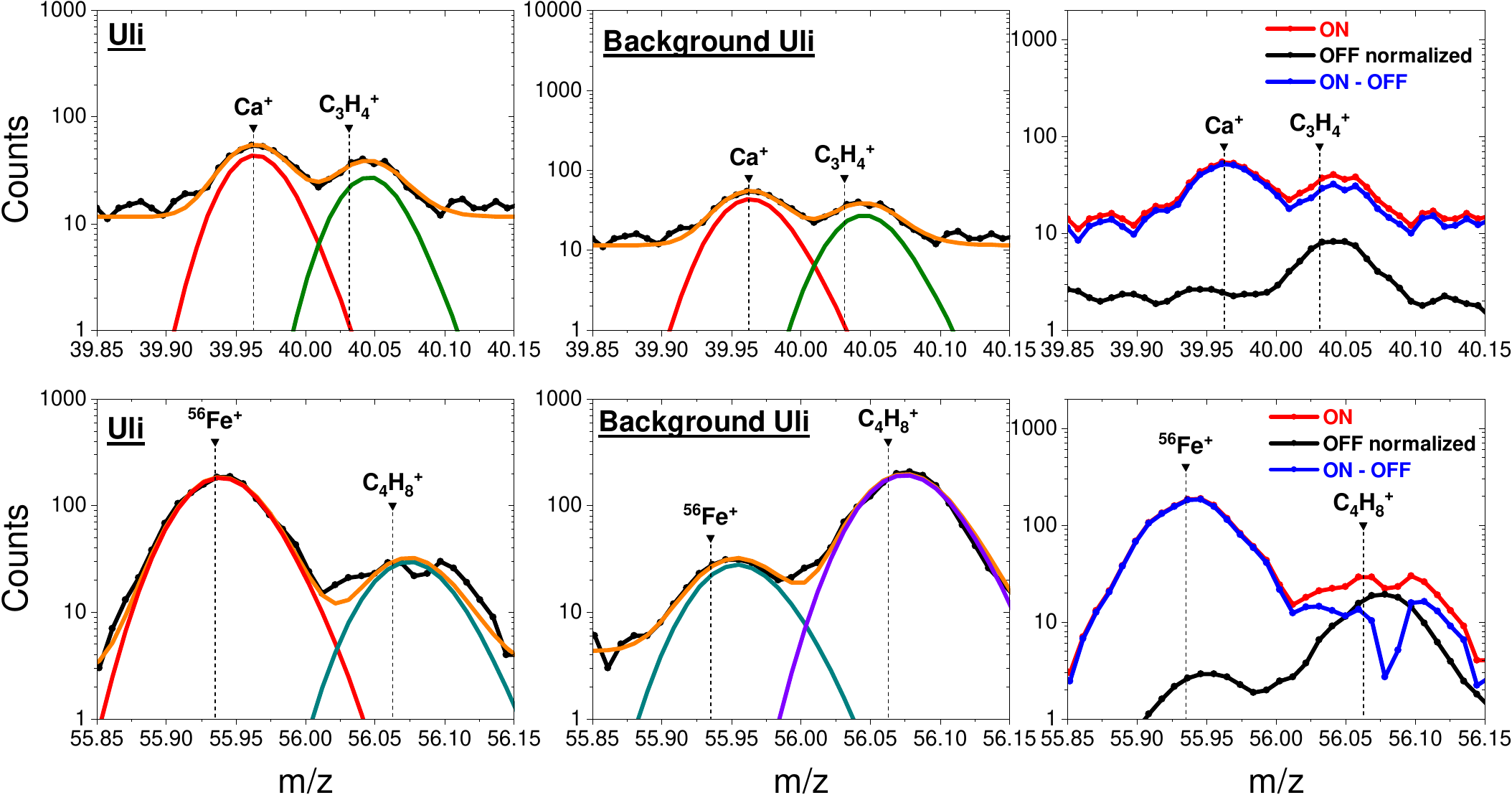}
    \caption{As Fig 1 showing the abundance of Ca and Fe on the particle Uli. The counts for Calcium are extremely low in comparison to the reference sample.}
    \label{FeCaParticles}
\end{figure*}